\documentstyle[12pt]{article}
\renewcommand{\a}{\alpha}
\newcommand{\DD}{{\bf D}}
\newcommand{\rot}{{\bf \rm rot}}
\newcommand{\be}{\begin{equation}}
\newcommand{\ee}{\end{equation}}
\newcommand{\ba}{\begin{eqnarray}}
\newcommand{\ea}{\end{eqnarray}}
\newcommand{\ul}{\underline}
\newcommand{\h}{\chi}
\renewcommand{\v}{\check}

\newcommand{\s}{\sigma}
\renewcommand{\d}{\partial}
\newcommand{\x}{{\bf x}}

\newcommand{\w}{\omega}

\newcommand{\e}{\epsilon}
\renewcommand{\Re}{{\rm Re\,}}
\renewcommand{\Im}{{\rm Im\,}}

\begin{document}
\begin{center}
{\Large \bf PAULI EQUATION AND THE METHOD OF SUPERSYMMETRIC
FACTORIZATION
}\\

\vspace{1cm} {\large M. V. Ioffe and A. I. Neelov}
\end{center}
\vspace{1cm} {\small Department of Theoretical Physics, University
of Sankt-Pe\-ters\-burg, 198504 Sankt-Pe\-ters\-burg, Russia.

E-mail: m.ioffe@pobox.spbu.ru and neelov@AN6090.spb.edu
\vspace{1cm}

{\bf Abstract.} We consider different variants of factorization of
a 2x2 matrix Schroe\-din\-ger/Pau\-li operator in two spatial
dimensions. They allow to relate its spectrum to the sum of
spectra of two scalar Schr\"odinger operators, in a manner similar
to one-dimensional Darboux transformations. We consider both the
case when such factorization is reduced to the ordinary
2-dimensional SUSY QM quasifactorization and a more general case
which involves covariant derivatives. The admissible classes of
electromagnetic fields are described and some illustrative
examples are given. } \vspace{1cm}
\section*{\large\bf \quad 1. Introduction}
In the ordinary one-dimensional supersymmetric quantum mechanics
(SUSY QM) \cite{witten}, a pair of Hamiltonians (superpartners) is
factorized as:
\ba
H^{(0)}&=&-\d^2+(\d\h)^2-\d^2\h= S^+S^-
\qquad \partial\equiv{d\over d x} \nonumber\\
H^{(1)}&=&-\d^2+(\d\h)^2+\d^2\h=S^-S^+\label{fac}
\ea
where
\ba
S^\pm\equiv \mp\d+\d\h;\qquad S^+=(S^-)^\dagger.
\ea
The arbitrary function $\h(x)$ is called a superpotential. The
(not necessarily normalizable) solution of the Schr\"odinger
equation for $H^{(0)}$ with zero energy is $\psi^{(0)}_0=e^{-\h}$.

In all cases when one has the factorization relations of the form
(\ref{fac}),
 the supercharge operators $Q^\pm$ and the superhamiltonian $H_S$ can be introduced \cite{witten}:
\ba
Q^+\equiv\left( \matrix{
                            0 & 0 \cr
                            S^- & 0
                                     } \right)\qquad
Q^-\equiv\left( \matrix{
                            0 & S^+ \cr
                            0 & 0
                                     } \right)=(Q^+)^\dagger\qquad
H_S\equiv\left( \matrix{
                            H^{(0)} & 0 \cr
                            0 & H^{(1)}
                                     } \right).\label{sudef}
\ea
The operators $S^\pm$ and $H^{(0)}$, $H^{(1)}$ are called the
components of the supercharge and of the superhamiltonian,
respectively. The relations of the SUSY algebra:
\ba
\{Q^+,Q^-\}=H_S\qquad (Q^+)^2=(Q^-)^2=0\qquad
[Q^\pm,H_S]=0\label{sualg}
\ea
follow from the definitions (\ref{sudef}) and the factorization
(\ref{fac}). The last equality in (\ref{sualg}) is equivalent to
the following intertwining relations:
\ba
H^{(1)}S^-=S^-H^{(0)}\qquad S^+H^{(1)}=H^{(0)}S^+.
\ea
The eigenfunctions of the Hamiltonians are interconnected by the
operators $S^\pm$:
\ba
\psi^{(0)}(x;E_{n}) = E_n^{-1/2}S^{+}\psi^{(1)} (x;E_{n}) \nonumber\\
\psi^{(1)} (x;E_{n})= E_n^{-1/2} S^{-}\psi^{(0)} (x;E_{n})
\nonumber
\ea
hence their eigenvalues coincide, except for the zero modes of the
supercharges.

Similarly to that, in two dimensions the following pair of
Hamiltonians can be factorized\footnote{It will be convenient to
denote the two-dimensional
vector with components $R_1, R_3$ as ${\bf R}$; $\s_i, i=1,...,3$
are the Pauli matrices. }:
\ba
H^{(0)}&=&-\Delta+(\d_i \h)^2-\s_3\Delta\h= S^+S^-\label{fac1}\\
H^{(1)}&=&-\Delta+(\d_i \h)^2+2\s_1\d_1\d_3\h+\s_3(\d_1^2-\d_3^2)\h=S^-S^+\label{fac2}\\
\d_i&\equiv&{\d\over\d x_i}\qquad (i=1,3)\qquad \Delta\equiv
\d_1^2+\d_3^2.\nonumber
\ea
This time, the matrix differential operators:
\ba
S^\pm=\mp\d_1-i\s_2\d_3+\s_1\partial_3\h +\s_3\partial_1\h \qquad
S^+=(S^-)^\dagger\label{S}
\ea
intertwine the Hamiltonians. The superpotential $\h(\x)$ is an
arbitrary real scalar function. Note that $H^{(0)}$ is a diagonal
Hamiltonian. The solution of its upper component with zero
energy is $\psi^{(0)}_0=e^{-\h}$.

 We remark that it would be natural to generalize the
1-dimensional factorization (\ref{fac}) onto the 3-dimensional
models, similarly to (\ref{fac1})-(\ref{S}). In this case
(\ref{S}) has to contain a term with $\d_2$ multiplied by some
Pauli matrix (in order to get a factorizable
Schroedinger-type Hamiltonian). However, all three Pauli matrices
have already been used in (\ref{S}), so a simple direct
three-dimensional generalization of (\ref{fac1})-(\ref{S}) is
impossible. Some generalizations for 3-dimensional case were
discussed in \cite{Vinet}, where $4\times 4$ Dirac matrices were
used, and in \cite{Rittenberg}. But the construction of the last
paper is not adapted to SUSY-diagonalization of Pauli operator.
The 3-dimensional factorization of the components of
superhamiltonian was considered in \cite{Levai} leading to
the Hamiltonian with terms linear in spatial derivatives (in
particular, as a spin-orbital coupling). In this case supercharges
with $O(3)-$rotational invariance were chosen. Here we generally
do not suppose any spatial symmetry of $S^{\pm}.$ Another possible
3-dimensional generalization of (\ref{fac1})-(\ref{S}) will be
discussed in the Section 7 of the present paper.

Returning back to the factorization relations (\ref{fac1}),(\ref{fac2})
one could again construct the supercharge operators and the
superhamiltonian\footnote{ In \cite{abis} a different definition
was used: the positions of the components ${\ul H}^{(0)}$ and
$H^{(1)}$ in $H_S$ were interchanged. For the readers familiar
with the two-dimensional SUSY QM \cite{abis},\cite{abif1} we may
add that this corresponds to choosing a different representation
of fermionic creation/annihilation operators. }:
\ba
Q^+\equiv\left( \matrix{
                            0 & 0 \cr
                            S^- & 0
                                     } \right)\qquad
Q^-\equiv\left( \matrix{
                            0 & S^+ \cr
                            0 & 0
                                     } \right)=(Q^+)^\dagger\nonumber\\
H_S= \left( \matrix{
                            H^{(0)} & 0 \cr
                            0 & H^{(1)}
                                     } \right)\equiv
\left( \matrix{
  {\ul{\ul H}}^{(0)}  &      0        & 0 \cr
      0               & {\ul H}^{(0)} & 0 \cr
      0               &      0        & H^{(1)}
                                     } \right)\label{sudef2}
\ea
where
\ba
{\ul{\ul H}}^{(0)}=-\Delta+(\d_i \h)^2-\Delta\h \qquad {\ul
H}^{(0)}=-\Delta+(\d_i \h)^2+\Delta\h\qquad i=1,3 \nonumber
\ea
and $H^{(1)}$ is still defined by the first equality of
(\ref{fac2}). Below we will call $H^{(0)}$ the diagonal component
of the superhamiltonian and $H^{(1)}$ the matrix one.

From the factorization relations (\ref{fac1}),(\ref{fac2}) it
follows again that
\ba
\{Q^+,Q^-\}=H_S;\qquad (Q^+)^2=(Q^-)^2=0\qquad
[Q^\pm,H_S]=0.\label{sualg2}
\ea

The last equality of (\ref{sualg2}) will again be equivalent to
the intertwining relations:
\ba
H^{(1)}S^-=S^-H^{(0)}\qquad S^+H^{(1)}=H^{(0)}S^+.\label{int}
\ea

The components of the supercharge operators (\ref{S}) can be
rewritten as:
\ba
S^-=
  \left( \matrix{
  Q^-_1 &  P^-_1 \cr
  Q^-_3 &  P^-_3
                  } \right)\qquad
S^+=
  \left( \matrix{
  Q^+_1 &  Q^+_3 \cr
  P^+_1 &  P^+_3
                  } \right)\label{connect}
\ea
where ($l,m=1,3$)
\ba
Q_l^{\pm}\equiv \mp\d_l+\partial_l\h\qquad
P_l^\pm\equiv\e_{lm}Q_m^\mp \qquad \e_{11}=\e_{33}=0\qquad
\e_{13}=-\e_{31}=1.\nonumber
\ea
If we plug (\ref{connect}) into the intertwining relations
(\ref{int}) we, after some algebra, will recover the usual
relations of the two-dimensional SUSY QM \cite{abis},\cite{abif1}:
\ba
H_{lm}^{(1)}Q_m^-&=&Q_l^-{\ul{\ul H}}^{(0)}\qquad
Q_l^+H_{lm}^{(1)}={\ul{\ul H}}^{(0)}Q_m^+\nonumber\\
H_{lm}^{(1)}P_m^-&=&P_l^-{\ul H}^{(0)}\qquad
P_l^+H_{lm}^{(1)}={\ul H}^{(0)}P_m^+\qquad l,m=1,3. \label{susy}
\ea

The eigenfunctions of the Hamiltonians $H^{(0)}$, $H^{(1)}$  are
interconnected by the operators $S^\pm$:
\ba
\psi^{(0)} (\x ;E_{n})= E_n^{-1/2} S^{+}\psi^{(1)} (\x ;E_{n}) \nonumber\\
\psi^{(1)} (\x ;E_{n})= E_n^{-1/2} S^{-}\psi^{(0)} (\x ;E_{n}).
\label{Spsi}
\ea
It follows that the eigenvalues of the Hamiltonians
$H^{(0)},H^{(1)} $ coincide, except for the zero modes of the
supercharges. If $H^{(1)}$ is not diagonalizable by constant
unitary transformations (see Section 3) then the above
supersymmetric transformations can simplify the derivation of its
spectrum by relating it to the spectrum of a diagonal Hamiltonian
$H^{(0)}$ \cite{abis},\cite{abif1}. In the most interesting
example  \cite{ai} one identifies $H^{(1)}$ with the Pauli
operator \cite{Ahi} of a non-relativistic spin $1/2$ particle in
external electromagnetic field.

The Pauli operator was studied in the framework of SUSY QM in
many papers (see \cite{Rittenberg}, \cite{Vinet},\cite{Clark} -
\cite{Nikitin2}). In these papers the Pauli operator was usually
identified with the {\bf total superhamiltonian}, while in the
present work the Pauli operator coincides with a {\bf component of
the superhamiltonian} (in this sense we continue the
approach of \cite{ai}). An advantage of our approach is
the opportunity to consider arbitrary values of the gyromagnetic
ratio $g=\frac{2\mu}{e}$ (in contrast to the usual restriction
that $g=2$). Also, we will consider the magnetic fields
having {\bf non-zero} components in the $(x_1, x_3)$ plane, for
which the Pauli problem is not diagonalizable by rotations. We
have to mention that interesting concrete models of Pauli operator
were investigated by means of SUSY QM: a particle with spin in the
field of monopole \cite{Vinet}, and a spin 1/2 particle in the
magnetic field of a thin straight wire with current
\cite{Voronin}. The SUSY-diagonalizability of Pauli operators
was considered in \cite{Nikitin1}, \cite{Nikitin2} but only for
external fields with definite spatial symmetries.

The paper is organized as follows. In Section 2 we consider
the simplest case, when the Pauli operator is literally identified
with the matrix component of the superhamiltonian . The Pauli
operator is originally three-dimensional but admits the separation
of the $x_2$ variable. We derive the expressions for the magnetic
field and the scalar potential in the Pauli operator and for the
superpotential that follow from such identification.

In Section 3 we enlarge the class of the Pauli operators, that
can be identified with the matrix component of the
superhamiltonian, by allowing
 constant unitary rotations, and describe the class of
superpotentials for which it is possible. The content of Sections
2 and 3 should be considered as a generalization of  \cite{ai} for
the case of nonzero electric current.

In Section 4 we investigate the case of a charged particle in
detail. It turns out that the separation of $x_2$ variable is not
trivial. The class of magnetic field for which the Pauli operator
of a charged particle can be identified (up to a unitary rotation)
with $H^{(1)}$ is described. It is smaller than the one for a
neutral particle.

In Section 5 we describe a new generalization of the SUSY QM
transformations. It involves the transition from the usual
derivatives to the covariant ones both in the components of the
supercharge operators and those of the superhamiltonian. We also
prove the general statement that the quasifactorization relations
\cite{abis},\cite{abif1} always describe a pair of factorized
Hamiltonians, one of which is diagonal.

In Section 6 we use these generalized transformations to
diagonalize the Pauli operators for more general configurations of
magnetic field in the case the gyromagnetic ratio is equal to 2.
In particular, the component $B_2$ of the magnetic field along the
$x_2$ direction can be arbitrary. Examples of the Pauli operators
allowing the above treatment are given in the end of Sections 2, 4
and 6.

In Section 7 we construct a three-dimensional
generalization of the two-di\-men\-sio\-nal factorization
relations described in two previous Sections.

\section*{\large\bf \quad 2.
Pauli Hamiltonian as a matrix component of the superhamiltonian. }
A non-relativistic particle with spin 1/2 in external
electromagnetic field is described \cite{Ahi} by the Pauli
Hamiltonian\footnote { From this moment on, ${\vec R} $ denotes a
three-dimensional vector with the components $R_{1}, R_{2},
R_{3}$.}:
\ba
 H_P = -{\vec D}^2-\mu{\vec\sigma}\cdot
  {\vec B}({\vec x}) +U({\vec x})
\label{hp1} \\
(\hbar=c=2m=1)\qquad \vec D\equiv \vec\partial-ie\vec A({\vec x})
\nonumber
\ea
where ${\vec A}({\vec x})$ is the electromagnetic vector
potential;
 ${\vec B} ({\vec x}) =\rot {\vec A}({\vec x})$ is the external magnetic field; $U({\vec x})$ is the external scalar potential
 (not necessarily an electrostatic one); $e, \mu$ are the charge and the
 magnetic moment\footnote{We stress that the value of the gyromagnetic
 ratio $g=2\mu/e$ can be {\bf arbitrary}.} of a particle.
 The operator (\ref{hp1}) acts on the two-component wave functions.

Let us assume that all external fields in (\ref{hp1}) depend on
two coordinates (for example $(x_1,x_3)\equiv {\bf x}$) only. Then
the wave function of a particle that moves freely along the $x_2$
direction can be written as:
 \ba
   \psi({\vec x}) = e^{-ikx_2}\psi({\bf x})\label{separ}
 \ea
and the Pauli Hamiltonian, acting on $\psi({\bf x})$, assumes the
form:
\ba
 H_P =-\DD^2+(k+eA_2({\bf x}))^2
 -\mu{\vec\sigma}\cdot {\vec B}({\bf x}) +U({\bf x}).  \label{hp2}
\ea

Insofar as the operator (\ref{hp2}) is a $2 \times 2$ matrix
differential
 operator in the two-dimensional space $(x_1, x_3)$, one can try to
identify it
 with the matrix component $H^{(1)}$ of the superhamiltonian:
\ba
H_P=H^{(1)}+E_0. \nonumber
\ea
This idea was set forth in \cite{ai}, but in that paper it was
additionally assumed that the magnetic field has no sources:
${\vec j}={\vec 0}$. In the present paper, on the contrary, we
consider the most general case of the equivalence of $H_P$ and
$H^{(1)}$, when the current density can be nonzero. Thus one can
try to consider a wider class of Pauli operators.

If such identification is possible, the corresponding Pauli
operator can be diagonalized by the supersymmetric transformations
from Section 1, and its spectrum is the sum of the spectra of
${\ul H}^{(0)}$ and ${\ul{\ul H}}^{(0)}$ (see (\ref{susy}),
(\ref{Spsi})). It is possible only if the following relations
between the external fields in the Pauli operator and the
superpotential $\chi({\bf x})$ are satisfied
:
  \ba
   -\mu {\vec B}&=&(2\partial_1\partial_3\chi,  0,
  (\partial_1^2-\partial_3^2)\chi) \label{bch} \\
   U({\bf x})&=&(\d_i \h)^2-(k+eA_2)^2+E_0\qquad i=1,3. \label{uch}
 \ea

 From the absence of first derivatives in $H^{(1)}$, and,
 consequently, in $H_P$, it follows that either $e=0$ (a neutral particle) or
 $A_1=A_3=0$.

 In the equation (\ref{uch}) all terms\footnote{ In the paper \cite{ai}, $A_2$ was chosen such that $k+eA_2$ is $k$-independent. We will not use this idea because
$A_2$ is a quantity that enters in the original three-dimensional
Pauli operator (\ref{hp1}) and therefore cannot depend on $k$. In
the following Sections
 we will discuss cases when $\chi$ depends on $k$, but for now we assume
 that it is $k$-independent.
}, except
 $(k+eA_2)^2$ and possibly $E_0$, have no
dependence on $k$.  The k-dependent parts of these two terms
cancel only if $e=0$ and $E_0=k^2$, therefore we assume this
choice of $e$ and $E_0$ in the remaining part of this Section.
From the Maxwell equation $\partial_i B_i=0$ and (\ref{bch}) one
can infer the following constraint onto the superpotential
$\chi({\bf x})$:  \ba
\partial_3(3\partial_1^2-\partial_3^2)\chi=0.  \label{div} \ea The general solution
of this equation has the form (see Appendix):  \ba \chi({\bf
x})=F(x_1)+G(-x_1/2+(\sqrt{3}/2)x_3)+H(-x_1/2-(\sqrt{3}/2)x_3)
   \label{FGH}
\ea
where $F, G, H$ are arbitrary thrice-differentiable
  functions of their arguments.

 The equalities (\ref{bch}),(\ref{uch}) allow us to express all the physical
 quantities in terms of these functions. In particular,
 \ba
   B_1({\bf x})=\frac{\sqrt{3}}{2\mu}(G''-H'') \qquad
   B_3({\bf x})=\frac{1}{\mu}(-F''+
   1/2(G''+H'')) \label{B13}
 \ea
and the electric current ${\vec j}({\bf x})= \rot {\vec B}({\bf
x})$ has, obviously, only one nonzero component:
\ba
j_2({\bf x})=  \frac{1}{4\pi\mu}[F'''+G'''+H''']\nonumber
\ea
The scalar potential (\ref{uch}) can then also be expressed in
terms of $F, G, H$:
\ba
U({\bf x})=
 F'^2+G'^2+H'^2-F'G'-F'H'-G'H'. \label{U}
\ea

 Thus the Pauli operator with the external fields as in (\ref{B13}),
(\ref{U}) coincides with the Hamiltonian $H^{(1)}$ and can be
diagonalized by the supersymmetric transformations. One can check
that in the case
 $j_2=0$ our solution for $\chi({\bf x})$ coincides with the one offered in
 \cite{ai} for the case of a neutral particle.

{\it Example }. Let us choose specific functions $F, G, H$ in
(\ref{FGH}):
\ba
  F(x)=G(x)=H(x)=ax^3 \nonumber
\ea
where $a$ is an arbitrary constant. Then, from (\ref{B13}),
\ba
   B_1=-\frac{9a}{\mu}x_3
   \qquad
   B_3=\frac{9a}{\mu}x_1.   \nonumber
\ea

The electric current that generates this system of external fields
is homogeneous and is directed along the axis $x_2$:
\ba
   j_2=-\frac{9a}{2\pi\mu}.\nonumber
\ea

The scalar potential (\ref{U}) in $H_P$  depends on $\rho^2\equiv
x_{1}^{2}+x_{3}^{2} $ only:
\ba
   U({\bf x})=(\partial_i\chi)^2=
\gamma\rho^4 \qquad \gamma=\frac{81a^2}{16} \nonumber
\ea
and the diagonal component of the superhamiltonian is proportional
to a unit matrix:
\ba
{\ul{\ul H}}^{(0)}= {\ul H}^{(0)}=
 -\Delta+\gamma\rho^4 \nonumber
\ea
since $\Delta\chi=0$ in this case. Thus the spectrum of the Pauli
operator in question consists of twofold degenerated levels of the
anharmonic oscillator $\gamma\rho^4$.

\section*{\large\bf \quad 3. Unitary rotation.}

In order to enlarge the class of external fields, for which the
Pauli Hamiltonian can be diagonalized by the supersymmetric
factorization, one can identify $H_P$ with the operator, which is
unitarily equivalent to the matrix component of the
superhamiltonian \cite{ai}:
\ba
 H_P=\widetilde H^{(1)}+E_0\qquad  \widetilde H^{(1)}= {\cal U}  H^{(1)} {\cal U}^+ \nonumber
\ea
where $ \cal{U} $ is a constant $2\times 2$ matrix $ {\cal U}
=\alpha_0+i{\vec \alpha}{\vec \sigma}$;\ \ $\alpha_0^2+{ \vec
\alpha}^2=1$; $ \alpha_0, \alpha_j $ are real constants; ${\vec
\sigma} =(\sigma_1,\sigma_2,\sigma_3) $.

Such rotation obviously doesn't change the spectrum of the matrix
operator $H^{(1)}$, but leads to additional freedom because of the
presence of the new parameters $ \alpha_0, \alpha_j $. Now the
relation between the components of the magnetic field and the
superpotential $\chi$ is more complicated compared to (\ref{bch}):
\ba
 \mu B_i=
[ (\alpha_0^2-{\vec \alpha}^2)\delta_{ik}+
 2(\alpha_j\alpha_0\epsilon_{ijk}+\alpha_k\alpha_i)]X_k\qquad
(i,k=1,2,3)   \label{bx}
\ea
where $ {\vec X}\equiv   (-2\partial_1\partial_3\chi, 0,
  (\partial_3^2-\partial_1^2)\chi).
$ The scalar potential is still defined by (\ref{uch}). Taking the
divergence of the equation (\ref{bx}), we obtain:
\ba
0=\mu\partial_iB_i=\partial_i [ (\alpha_0^2-{\vec
\alpha}^2)\delta_{ik}+
 2(\alpha_j\alpha_0\epsilon_{ijk}+\alpha_k\alpha_i)]X_k =\nonumber\\=
-(\alpha_0^2+\alpha_1^2-\alpha_2^2-\alpha_3^2)2\partial_1^2\partial_3\chi+
 (\alpha_0^2-\alpha_1^2-\alpha_2^2+\alpha_3^2)
 (\partial_3^3-\partial_1^2\partial_3)\chi+ \nonumber\\
+(2\alpha_1\alpha_3-2\alpha_2\alpha_0)
 (\partial_1\partial_3^2-\partial_1^3)\chi-
 (2\alpha_2\alpha_0+2\alpha_1\alpha_3)
 2\partial_1\partial_3^2\chi.\nonumber
\ea

Hence, the superpotential $\chi$ satisfies the following partial
differential equation with constant coefficients:
\ba
 \biggl[ 2(\alpha_2\alpha_0-\alpha_1\alpha_3)\partial_1^3-
 (3\alpha_0^2+\alpha_1^2-3\alpha_2^2-\alpha_3^2)\partial_1^2\partial_3-
 2(\alpha_1\alpha_3+3\alpha_2\alpha_0)\partial_1\partial_3^2+      \label{pde}        \\
 +(\alpha_0^2-\alpha_1^2-\alpha_2^2+\alpha_3^2)\partial_3^3\biggr] \chi=0.  \nonumber
\ea
The equation (\ref{div}) is a partial case of (\ref{pde}) with
$\alpha_0=1; \alpha_i=0$.

It is shown in the Appendix that the superpotentials $\chi$,
satisfying this equation and therefore allowing the identification
$H_P=\widetilde H^{(1)}+E_0$, must have the form\footnote{If some
of $t_i$ are complex one can take only such functions $F,G,H$ that
the superpotential is a real function.}:
\ba
\chi=F(t_1x_3+x_1)+G(t_2x_3+x_1)+H(t_3x_3+x_1) \label{FG1}
\ea
where $t_1, t_2, t_3$ are the roots of the characteristic
polynomial, which is determined by the operator in (\ref{pde}):
\ba
 (2(\alpha_2\alpha_0-\alpha_1\alpha_3)\partial_1^3-
 (3\alpha_0^2+\alpha_1^2-3\alpha_2^2-\alpha_3^2)\partial_1^2\partial_3-
 2(\alpha_1\alpha_3+3\alpha_2\alpha_0)\partial_1\partial_3^2+           \nonumber \\
 +2( \alpha_0^2-\alpha_1^2-\alpha_2^2+\alpha_3^2)\partial_3^3)=
 c_0(\partial_3-t_1\partial_1)(\partial_3-t_2\partial_1)(\partial_3-t_3\partial_1)
\label{roots}\\
c_0=2( \alpha_0^2-\alpha_1^2-\alpha_2^2+\alpha_3^2). \nonumber
\ea

In the case of a neutral particle ($e=0$ but $\mu\ne 0$) any
choice of $\alpha_i$ and the corresponding superpotential
(\ref{FG1}), leads to a Pauli operator determined by
(\ref{hp2}),(\ref{uch}),(\ref{bx}), that is  diagonalizable by
supersymmetric transformations.

For a charged particle ($e\ne 0$) there are stronger restrictions:
as we mentioned in Section 2, this approach is valid only for
$A_1=A_3=0$, and, consequently, $B_2=0$ and $j_1=j_3=0$. Then,
from (\ref{bx}),
\ba
(2\alpha_3\alpha_0+2\alpha_1\alpha_2)B_1({\bf x})+
(2\alpha_3\alpha_2-2\alpha_1\alpha_0)B_3({\bf x})=0.
\label{BaBa}
\ea
If either of the constants in (\ref{BaBa}) is nonzero, the Pauli
operator can be trivially diagonalized by a constant unitary
rotation, after which $\vec B$ is directed along $x_3$. Hence, the
interesting situation is when
$2\alpha_3\alpha_0+2\alpha_1\alpha_2=2\alpha_3\alpha_2-2\alpha_1\alpha_0=0$.
One can check that it is possible only in the following two cases:

1) $\alpha_3=\alpha_1=0$

2) $\alpha_2=\alpha_0=0$

Unfortunately, in the variant 1) we couldn't find solutions that
satisfy (\ref{uch}) with the scalar and vector potentials $U$ and
${\vec A}$ independent on $k$.

In the variant 2) the roots $t_1, t_2, t_3$ (\ref{roots}) have the
form

\ba
 t_1=\frac{2\alpha_1\alpha_3}{\alpha_3^2-\alpha_1^2} \qquad
 t_2=i\qquad t_3=-i.\nonumber
\ea
Hence, taking the reality of the superpotential into account,
(\ref{FG1}) leads to
\ba
\chi=F\biggl(x_1+\frac{2\alpha_1\alpha_3}{\alpha_3^2-\alpha_1^2}x_3\biggr)+
    \Re g(z)\qquad z=x_1+ix_3     \label{FG3}
\ea
where $g(z)=2G(z)$ is an arbitrary holomorphic function of $z$.

\section*{\large\bf \quad 4. The dependence of the spectrum on the
momentum along the axis $x_2$.}

It can be shown that in the case of a charged particle ($e\ne 0$)
the superpotential (\ref{FG3}) can satisfy (\ref{uch}) with the
scalar and vector potentials $U$ and ${\vec A}=(0,A_2,0)$ {\it
independent on} $k$ only if $F=0$:
\ba
\chi=\Re g(z).\label{chig}
\ea
Then (\ref{bx}) has the following solution\footnote{ A similar
expression for the magnetic field was found in \cite{ai}. However
in the present text we allow
$j_2=\partial_1B_3-\partial_3B_1\nonumber$ to be nonzero (other
two components of the current density are zero because of our
assumptions). This is possible if $g_0(z)$ has singularities,
i.e., it is not an entire function (see Example below). Another
difference of the approach of \cite{ai} is that it did not allow
to obtain the dependence of energy levels on $k$, in contrast to
ours. }:
\ba
B_1=-{2\over\mu}\Im g''\qquad B_3=-{2\over\mu}\Re g''\qquad
A_2=-{2\over\mu}\Re g'+C.\label{AgC}
\ea
Let us choose
\ba
g(z)=g_0(z)-{2e\over\mu}kz \label{gg0}\\
C=C_0-{4e\over\mu^2}k \nonumber
\ea
 where both
$g_0$ and $C_0$ do not depend on $k$. Then from (\ref{AgC})
\ba
B_1=-{2\over\mu}\Im g_0''\qquad B_3=-{2\over\mu}\Re g_0'' \qquad
A_2=-{2\over\mu}\Re g_0'+C_0.\label{Ag0}\ea

Note that $A_2$ doesn't depend on $k$. Now let us prove that $U$
can also be made $k$-independent. Plugging
(\ref{chig}),(\ref{gg0}),(\ref{Ag0}) into (\ref{uch}), we can
check that
\ba
U=|g_0'|^2-\biggl({2e\over\mu}\Re
g_0'-C_0e\biggr)^2+E_0-\biggl(1-{4e^2\over\mu^2}\biggr)k^2-2C_0ek.\label{UE}
\ea

By choosing
\ba
E_0=(1-{4e^2\over\mu^2})k^2+2C_0ek \label{Eg0}
\ea
we can make $U$ $k$-independent. Indeed, from (\ref{UE})
\ba
U=|g_0'|^2-\biggl({2e\over\mu}\Re g_0'-C_0e\biggr)^2.\label{Ug0}
\ea

Plugging (\ref{chig}),(\ref{gg0}) into (\ref{fac1}), we see that
the Pauli operator with external fields (\ref{Ag0}),(\ref{Ug0}) is
equivalent to the diagonal Hamiltonian
\ba
H^{(0)}=-\Delta+\biggl|g_0'-{2e\over\mu}k\biggr|^2-\s_3\Delta\Re
g_0.\label{H0g0}
\ea
Note that $k$ enters into $H^{(0)}$ nontrivially.  The last term
in $H^{(0)}$ is zero everywhere except the points where $g$ has
singularities.

{\it Example .} Let us choose a specific function $g_0(z)$ in
(\ref{gg0}):
\ba
g_0(z)=a\ln z\nonumber
\ea
a being a real constant. Then we readily obtain from the
corresponding general expressions that
\ba
A_2=-{2ax_1\over \mu\rho^2}+C_0\qquad
B_1=-{4a\over\mu}{x_1x_3\over \rho^4}\qquad
B_3={2a\over\mu}{x_1^2-x_3^2\over \rho^4}\nonumber\\
j_2=-{a\over\mu}\partial_1\delta(x_1)\delta(x_3).\nonumber
\ea

This electric current density can be interpreted as a "dipole
current" - two thin linear currents in converse directions, the
distance between them being small.

In this case, (\ref{Ug0}) turns into:
\ba
 U({\bf x})=\frac{a^2}{\rho^2}-\biggl(\frac{aex_1}{\mu\rho^2}-C_0e\biggr)^2.\nonumber
\ea

From (\ref{H0g0}) it follows that the spectrum of our Pauli
operator coincides with the spectrum of the diagonal Hamiltonian
$H^{(0)}$, which has the form:
\ba
H^{(0)}=-{\bf\partial}^2+\biggl|\frac{a}{z}-{2ek\over\mu}\biggr|^2-2\pi
a\delta(x_1)\delta(x_3).\nonumber
\ea

\section*{\large\bf \quad 5. More general factorization method.}

In the Pauli operator (\ref{hp2}) each derivative enters in the
covariant derivative combination $D_j\equiv \d_j-ieA_j$.
Therefore, it seems natural to  use the covariant derivatives in
the operators (\ref{fac1})-(\ref{S}) too\footnote{Note that in the
simplest particular case $\chi=0$ the operators (\ref{SD}) can be
reduced to the SUSY generators given in \cite{Rittenberg},
 \cite{Clark} by the multiplication onto $\s_1$
or $\s_3$ and a redefinition of axes.  }:
\ba
S^\pm=\mp D_1-i\s_2D_3+\s_1\partial_3\h +\s_3\partial_1\h \qquad
S^+=(S^-)^\dagger.\label{SD}
\ea

The so defined intertwining operators generate the following pair
of factorized Hamiltonians:
\ba H^{(0)}&=&
S^+S^-=-\DD^2+(\d_i \h)^2-e\s_2B_2-\s_3\Delta\h \label{facD1}\\
H^{(1)}&=&S^-S^+=-\DD^2+(\d_i \h)^2+e\s_2B_2+2\s_1\d_1\d_3\h+
\s_3(\d_1^2-\d_3^2)\h \label{facD2}\\
B_2&\equiv& \d_3A_1-\d_1A_3\qquad i=1,3.\nonumber
\ea

The Hamiltonians (\ref{fac2}) obey the intertwining equations
similar to (\ref{int}):
\ba
H^{(1)}S^-=S^-H^{(0)}\qquad S^+H^{(1)}=H^{(0)}S^+.\nonumber
\ea
and their eigenfunctions are interconnected.

Similarly to Sections 2,3 one can try to identify $H^{(1)}$ with a
Pauli operator up to constant unitary rotations:
\ba
   H_P= \widetilde H^{(1)} +E_0\qquad
   \widetilde H^{(1)}={\cal U}  H^{(1)} {\cal U}^+ \label{trnD}
\ea
and $H^{(0)}$ is unitary equivalent to a diagonal Hamiltonian:
\ba
  H^{(0)}= {\cal V}^+  \widetilde H^{(0)}{\cal V} \label{h0}\\
{\widetilde H}^{(0)}= \left( \matrix{
  {\ul{\ul {\widetilde H}}}^{(0)} & 0 \cr
  0 & {\ul {\widetilde H}}^{(0)}
                  } \right).\label{diag}
\ea
If such identification is possible, the corresponding Pauli
operator can be diagonalized by the above supersymmetric-like
transformations and its spectrum is the sum of the spectra of
${\widetilde{\ul H}}^{(0)}$ and ${\widetilde{\ul{\ul H}}}^{(0)}$
Below we shall show that the more general factorization method
presented in this Section allows us to diagonalize a broader class
of Pauli operators than the usual supersymmetric methods described
in the previous Sections.

From (\ref{facD1})-(\ref{h0}) one may infer the new factorization
relations:
\ba
{\widetilde H}^{(0)}= {\widetilde S}^+{\widetilde S}^-\qquad
{\widetilde H}^{(1)}={\widetilde S}^-{\widetilde S}^+ \qquad
{\widetilde S}^-\equiv \cal{U}S^-\cal{V}^+\qquad {\widetilde
S}^+\equiv \cal{V}S^+\cal{U}^+=  ({\widetilde S}^-)^\dagger.
\label{factil}
\ea
Note that from (\ref{factil}) we can construct the SUSY algebra
relations, similar to (\ref{sudef}),(\ref{sualg}), or
(\ref{sudef2}),(\ref{sualg2}), and thus obtain a new realization
of SUSY QM in two dimensions, that differs from the one given in
\cite{abis},\cite{abif1}.

From (\ref{factil}) the following intertwining relations can be
derived:
\ba
{\widetilde H}^{(1)} {\widetilde S}^-={\widetilde S}^-{\widetilde
H}^{(0)}\qquad {\widetilde S}^+{\widetilde H}^{(1)}= {\widetilde
H}^{(0)} {\widetilde S}^+\nonumber
\ea
and the eigenfunctions of ${\widetilde H}^{(0)}$, ${\widetilde
H}^{(1)}$ are interconnected by the operators ${\widetilde
S}^\pm$:
\ba
{\widetilde \psi}^{(0)} (\x ;E_{n})= E_n^{-1/2} {\widetilde S}^{+}{\widetilde \psi}^{(1)} (\x ;E_{n}) \nonumber\\
{\widetilde \psi}^{(1)} (\x ;E_{n})= E_n^{-1/2} {\widetilde
S}^{-}{\widetilde \psi^{(0)} } (\x ;E_{n}). \nonumber
\ea
Thus  the eigenvalues of the Hamiltonians ${\widetilde H}^{(0)}$,
${\widetilde H}^{(1)}$ again coincide up to the zero modes of the
supercharge components ${\widetilde S}^{\pm}$.

Now let us parameterize the matrices ${\widetilde S}^\pm$ in the
following way, which is similar to (\ref{connect}):
\ba
{\widetilde S}^-=
  \left( \matrix{
  {\widetilde Q}^-_1 &  {\widetilde P}^-_1 \cr
  {\widetilde Q}^-_3 &  {\widetilde P}^-_3
                  } \right)\qquad
{\widetilde S}^+=
  \left( \matrix{
  {\widetilde Q}^+_1 &  {\widetilde Q}^+_3 \cr
  {\widetilde P}^+_1 &  {\widetilde P}^+_3
                  } \right).\label{def}
\ea
Plugging these operators into (\ref{factil}),(\ref{diag}),  one
can infer the following quasifactorization relations:
\ba
{\ul {\ul {\widetilde H}}}^{(0)}={\widetilde Q}_l^+{\widetilde
Q}_l^-\qquad {\ul {\ul {\widetilde H}}}^{(0)}={\widetilde
P}_l^+{\widetilde P}_l^-\qquad {\widetilde
H}_{lm}^{(1)}={\widetilde Q}_l^-{\widetilde Q}_m^+
+{\widetilde P}_l^-{\widetilde P}_m^+\label{quasyfac}\\
{\widetilde P}_l^+{\widetilde Q}_l^-={\widetilde Q}_l^+{\widetilde
P}_l^-=0 \qquad l,m=1,3.\nonumber
\ea
From them we can deduce an analog of the intertwining relations
(\ref{susy}):
\ba
{\widetilde H}_{lm}^{(1)}{\widetilde Q}_m^-&=& {\widetilde
Q}_l^-{\ul{\ul {\widetilde H}}}^{(0)}\qquad {\widetilde
Q}_l^+{\widetilde H}_{lm}^{(1)}=
{\ul{\ul {\widetilde H}}}^{(0)}{\widetilde Q}_m^+\nonumber\\
{\widetilde H}_{lm}^{(1)}{\widetilde P}_m^-&=& {\widetilde
P}_l^-{\ul {\widetilde H}}^{(0)}\qquad {\widetilde
P}_l^+{\widetilde H}_{lm}^{(1)}= {\ul {\widetilde
H}}^{(0)}{\widetilde P}_m^+\qquad l,m=1,3. \label{gensusy}
\ea
One may also note that ${\widetilde H}^{(1)}$  can be decomposed
as:
\ba
{\widetilde H}_{lm}^{(1)}= {\ul {\widetilde
H}}_{lm}^{(1)}+{\ul{\ul {\widetilde H}}}_{lm}^{(1)}\qquad {\ul
{\widetilde H}}^{(1)}_{lm}\equiv {\widetilde Q}_l^-{\widetilde
Q}_m^+\qquad {\ul{\ul {\widetilde H}}}^{(1)}_{lm}\equiv
{\widetilde P}_l^-{\widetilde P}_m^+
\ea
such that ${\ul {\widetilde H}}_{lm}^{(1)}{\ul{\ul {\widetilde
H}}}_{mn}^{(1)}= {\ul{\ul {\widetilde H}}}_{lm}^{(1)}{\ul
{\widetilde H}}_{mn}^{(1)}=0$, and the relations (\ref{gensusy})
can be rewritten in the form:
\ba
{\ul {\widetilde H}}_{lm}^{(1)}{\widetilde Q}_m^-&=& {\widetilde
Q}_l^-{\ul{\ul {\widetilde H}}}^{(0)}\qquad {\widetilde Q}_l^+{\ul
{\widetilde H}}_{lm}^{(1)}=
{\ul{\ul {\widetilde H}}}^{(0)}{\widetilde Q}_m^+\nonumber\\
{\ul{\ul {\widetilde H}}}_{lm}^{(1)}{\widetilde P}_m^-&=&
{\widetilde P}_l^-{\ul {\widetilde H}}^{(0)}\qquad {\widetilde
P}_l^+{\ul{\ul {\widetilde H}}}_{lm}^{(1)}= {\ul {\widetilde
H}}^{(0)}{\widetilde P}_m^+\qquad l,m=1,3. \label{decomp}
\ea
The relations (\ref{quasyfac})-(\ref{decomp}) have exactly the
same form as the well known SUSY QM quasifactorization relations
and the SUSY QM intertwining relations \cite{abis},\cite{abif1}.
The former are reduced to the latter in case ${\cal U}={\cal V}=I$
and $A_1=A_3=0$.

From the above algebra we see that the {\it any}
quasifactorization relations  in fact describe a pair of {\it
factorized} (in the sense of (\ref{factil})) Hamiltonians, one of
which is diagonal. This is a rather general statement, because the
relations (\ref{quasyfac})-(\ref{decomp}) follow from the
definitions (\ref{diag})-(\ref{def}), no matter what form the
components of the supercharge ${\widetilde Q}_l^\pm,{\widetilde
P}_l^\pm$ have.

\section*{\large\bf \quad 6. The cases when $H^{(0)}$ is diagonalizable.}

Eq. (\ref{trnD}) is equivalent to the following pair of equations:
\ba
 \mu B_i&=&
[ (\alpha_0^2-{\vec \alpha}^2)\delta_{ik}+
 2(\alpha_j\alpha_0\epsilon_{ijk}+\alpha_k\alpha_i)]Y_k \qquad (i,k=1,2,3)
 \label{BBX}\\
   U({\bf x})&=&({\bf \partial}\chi)^2-(k+eA_2)^2+E_0 \label{UBA}
 \ea
where $ {\vec Y}\equiv   (-2\partial_1\partial_3\chi,-eB_2,
  -(\partial_1^2-\partial_3^2)\chi).
$

Let us consider for simplicity the case when the gyromagnetic
ratio $2\mu/e$ is equal to\footnote{This is the first
instance when we fix the gyromagnetic ratio to the same value
$g=2$ that appeared in \cite{Rittenberg}, \cite{Clark}.
However, in this text we consider a much larger
class of the configurations of external fields compared to
\cite{Rittenberg}, \cite{Clark}, because, in contrast to them, we
allow the magnetic field component in the $(x_1,x_3)$ plane to be
nonzero.} $g=2$ and $\a_0=\a_2=0$. Then (\ref{BBX}) is reduced
to (\ref{bx}), which is solved in the Section 3, the solution
being given by (\ref{FG1}). Assume further that $\alpha_3=1;\
\a_1=0$, i.e., ${\cal U}=\s_3$. Then, Eq. (\ref{FG1}) is reduced
to a partial case of (\ref{FG3}):
\ba
\chi=F(x_1)+
    \Re g(z)\qquad z=x_1+ix_3.      \label{F1g}
\ea

Plugging the superpotential (\ref{F1g}) into (\ref{BBX}) we
obtain:
\ba
 B_1=-{2\over e}\Im g''\qquad
 B_3=-{2\over e}\Re g''-{1\over e}F''\qquad A_2=-{2\over e}\Re g'-{1\over e}F'+const.\nonumber
\ea
The first component $H^{(0)}$ of the superhamiltonian is unitarily
equivalent to a diagonal Hamiltonian in the following
two\footnote{We do not consider the  case when $A_1=A_3=0$ that
has already been investigated above.} cases:

a) there is some constant $C_B$ such that $B_2=C_B\Delta\h$

b) $\Delta\h=0$.

Unfortunately, in the case a) it is impossible to satisfy
(\ref{UBA}) when both $U$ and $A_2$ are $k$-independent. In the
rest of this Section we will consider the case b). Then the
Hamiltonian $H^{(0)}$ is already diagonal; therefore, in the rest
of the article we will assume that ${\cal V}=I$.

If the superpotential $\chi$ (\ref{F1g}) satisfies the condition
$\Delta\h=0$, it follows that $F(x_1)$ is a linear function. One
can check that we can set $F(x_1)=0$ without narrowing the class
of Pauli operators that can be diagonalized by our factorization
method. So,
\ba
\chi=\Re g(z).   \label{chige}
\ea

This is the same superpotential as in (\ref{chig}). However, this
time $g(z)$ cannot contain singularities anywhere except at the
infinity, because otherwise $\Delta\h$ would be nonzero.

The magnetic field components $B_1,B_3$ are obtained from the same
equation as (\ref{bx}),  therefore they will have the
form\footnote{ We keep taking into account that the gyromagnetic
ratio $2\mu/e$ is fixed at 2.} (\ref{AgC}):
\ba
B_1=-{2\over e}\Im g''\qquad B_3=-{2\over e}\Re g''\qquad
A_2=-{2\over e}\Re g'+C.\nonumber
\ea
 The component $B_2$ can be arbitrary.
Similarly to the Section 4, we choose the superpotential and the
constant in the magnetic potential in the form  (\ref{gg0}):
\ba
g(z)=g_0(z)-2kz \label{gg2} \\
C=C_0-{4\over e^2}k  \nonumber
\ea
Then the magnetic field components $B_1, B_3$ take the form
(\ref{Ag0}):
\ba
  B_1=-{2\over e}\Im g_0''\qquad B_3=-{2\over e}\Re g_0''  \label{BB}
\ea
while $B_2$ remains arbitrary. We get the scalar field $U(\x)$
from (\ref{UBA}), which coincides with (\ref{uch}). If we assume
that $E_0$ has the form (\ref{Eg0}):
\ba
E_0=-3k^2+2C_0ek \nonumber
\ea
then  the scalar  field will be defined by (\ref{Ug0}):
\ba
U=|g_0'|^2-\biggl({2}\Re g_0'-C_0e\biggr)^2\nonumber
\ea
Thus, same as in the Section 4, all physical fields in the Pauli
operator are $k$-independent.

Plugging (\ref{chige}),(\ref{gg2}) into (\ref{facD1}), we see that
the spectrum of the Pauli operator coincides with the spectrum of
the Hamiltonian
\ba
H^{(0)}=-\DD^2-\s_2B_2+\bigl|g_0'-2k\bigr|^2.\nonumber
\ea
After a unitary rotation by ${\cal V}=(\s_2+\s_3)/\sqrt{2}$,
$H^{(0)}$ will turn into a diagonal Hamiltonian analogous to
(\ref{H0g0}):
\ba
\widetilde
H^{(0)}=-\DD^2-\s_3B_2+\bigl|g_0'-2k\bigr|^2.\label{Htil}
\ea

{\it Example .} Let us choose a specific function $g(z)$ in
(\ref{chige}):
\ba
g(z)={\w\over 6}z^3.\nonumber
\ea
Then we readily obtain from the corresponding general
 expressions (\ref{BB})-(\ref{Htil}) that
$$
{\vec B}=-{2\w\over e}(x_3,0,x_1)+(0,B_2,0) \qquad {\vec j}={\vec
0}
$$
$$
U={\w^2\over 4}\rho^4-(\w(x_1^2-x_3^2)+const)^2\qquad \widetilde
H^{(0)}=-\DD^2-\s_3B_2+\biggl|{\w\over 2}z^2-2k\biggr|^2.
$$

\section*{\large\bf \quad 7. A three-dimensional factorization.}
In this Section we will show that the above factorization
method allows a three-dimensional generalization. It is realized
by the operators
\ba
R^\pm\equiv \mp{\vec\s}\cdot{\vec D}+V_0(\vec x);\qquad
R^+=(R^-)^\dagger\label{Rpm}
\ea
that can be used  for the factorization\footnote{The factorization
(\ref{Rpm}) - (\ref{Hv1}) was proposed in \cite{cooper2} for the
study of massless Dirac operators in Euclidean spacetime.} of the
following pair of Hamiltonians:
\ba
\v H^{(0)}&\equiv & R^+R^-=-\vec D^2-\vec\s\cdot
(\vec\d V_0+e\vec B)+V_0^2;\label{Hv0}\\
\v H^{(1)}&\equiv & R^-R^+=-\vec D^2+\vec\s\cdot (\vec\d V_0-e\vec
B)+V_0^2,\label{Hv1}
\ea
where again $\vec D=\vec\d-ie\vec A;\ \ \vec B=\rot \vec A$;\ \
$V_0(\vec x)$ is an arbitrary real scalar function.

The operators (\ref{Hv0}), (\ref{Hv1}) can then be treated in the
same way as in the Section 5: $\v H^{(0)}$ can be made unitarily
equivalent to a Pauli operator in three dimensions (\ref{hp1}):
\ba
   H_P= \widetilde {\v H}^{(1)} +E_0\qquad
   \widetilde {\v H}^{(1)}=\v{\cal  U}  \v H^{(1)} \v{\cal
   U}^+,\label{UHU}
\ea
and $\v H^{(0)}$ to a diagonal Hamiltonian $\widetilde{\v
H}^{(0)}$:
\ba
  \v H^{(0)}= {\v{\cal V}}^+  \widetilde{\v H}^{(0)}{\v{\cal V}}.\label{VHV}
\ea

Now let us prove that the operators (\ref{Rpm}) - (\ref{Hv1}) are
indeed a generalization of (\ref{SD}) - (\ref{facD2}). Namely,
assume that $V_0$ and $\vec A$ depend on two coordinates
$(x_1,x_3)$ only. Then, similarly to Section 2, the wave function
of a particle along the $x_2$ direction has the form
(\ref{separ}), and the operators (\ref{Rpm}) - (\ref{Hv1}), acting
on $\psi(\x)$, assume the form:
\ba
R^\pm\to \hat R^\pm&=&
\mp\biggl[\s_1D_1+\s_3D_3-i(k+eA_2)\s_2\biggr]+V_0;\label{hatR}\\
\v H^{(0)}\to\hat H^{(0)}&=&-\DD^2+(k+eA_2)^2-\vec\s\cdot
(\vec\d
V_0+e\vec B)+V_0^2=\hat R^+\hat R^-;\label{hatH0}\\
\v H^{(1)}\to\hat H^{(1)}&=&-\DD^2+(k+eA_2)^2+\vec\s\cdot(\vec\d
V_0-e\vec B)+V_0^2=\hat R^-\hat R^+.\label{hatH1}
\ea
The Pauli operator and the diagonal Hamiltonian in (\ref{UHU}),
(\ref{VHV}) will also transform to two-dimensional operators.

 Now, if we assume that $V_0=\d_3\chi$ and $A_2={1\over
e}(\d_1\chi-k)$ then one can check that (\ref{hatR}) -
(\ref{hatH1}) can be rewritten in terms of the operators from the
Section 5:
\ba
\hat R^+=\s_1S^+;\qquad  \hat R^-=S^-\s_1;\qquad
 \hat H^{(0)}= H^{(0)};\qquad  \hat H^{(1)}=
\s_1 H^{(1)}\s_1\nonumber
\ea
where $S^\pm$ are defined in (\ref{SD}), and $\widetilde H^{(0)}$,
$\widetilde H^{(1)}$ in (\ref{facD1}), (\ref{facD2}). Then one can
check that the two-dimensional versions of (\ref{UHU}),
(\ref{VHV}) are equivalent to (\ref{trnD}), (\ref{h0}), if we
set\footnote{Note that the choice of the unitary matrices
${\cal V}={I};\ \ {\cal U}=\s_2$, that we have widely used in
Sections 3-6, is thus equivalent to the simplest choice in the new
formalism of this Section: $\v{\cal V}=\v{\cal U}=I$.}:
\ba
{\v{\cal U}}={{\cal U}}\s_1;\qquad {\v{\cal V}}={{\cal V}}.
\nonumber
\ea

 Thus, we have proved that the operators (\ref{Rpm}) - (\ref{Hv1})
are indeed a generalization of (\ref{SD}) - (\ref{facD2}). Such
generalization may allow one to diagonalize a more general
class of three-dimensional Pauli operators.

\section*{\normalsize\bf Acknowledgements}

This work has been partially supported by grant of Russian
Foundation of Basic Researches (N  02-01-00499).
\section*{\large\bf \quad Appendix.}
It is straightforward that any differential operator of the form
\ba
c_0\partial_3^3+c_1\partial_3^2\partial_1+c_2\partial_3\partial_1^2+c_3\partial_1^3
\nonumber
\ea
with $c_{0} \ne 0$ can be factorized into a product of
differential operators of first order:
\ba
 c_0(\partial_3-t_1\partial_1)(\partial_3-t_2\partial_1)
 (\partial_3-t_3\partial_1) \nonumber
\ea
where $t_1, t_2, t_3$ are the roots of the polynomial
$c_0x^3+c_1x^2+c_2x+c_3$. Assume for simplicity that these roots
are different. Then the
 following statement is true: if
\ba
 (\partial_3-t_1\partial_1)(\partial_3-t_2\partial_1)
 (\partial_3-t_3\partial_1)f=0   \label{stm}
\ea
then $
 f=F(t_1x_3+x_1)+G(t_2x_3+x_1)+H(t_3x_3+x_1).
$

\underline{Proof}: if (\ref{stm}) is satisfied, then there exists
a function $f_0(t_1x_3+x_1)$ such that
\ba
 (\partial_3-t_2\partial_1)(\partial_3-t_3\partial_1)f=
 f_0(t_1x_3+x_1).  \label{ff0}
\ea
Insofar as $t_1, t_2, t_3$ are assumed to be different, one can
show:
 there exists a function $F(t_1x_3+x_1)$ (see below) such that (\ref{ff0})
 is equivalent to the following equation:
\ba
 (\partial_3-t_2\partial_1)(\partial_3-t_3\partial_1)(f-F(t_1x_3+x_1))=0
  \label{df0}
\ea
where $(\partial_3-t_2\partial_1)(\partial_3-t_3\partial_1)
F(t_1x_3+x_1) =
 f_0(t_1x_3+x_1).$

Similarly, from (\ref{df0}) it follows that there exists a
function $G(t_2x_3+x_1) $ such that
$$(\partial_3-t_3\partial_1)\biggl(f-F(t_1x_3+x_1)-G(t_2x_3+x_1)\biggr) = 0.$$
Therefore, there is $ H(t_3x_3+x_1)$, such that
\ba
   f-F(t_1x_3+x_1)-G(t_2x_3+x_1)-H(t_3x_3+x_1)=0.\nonumber
\ea
Q.E.D.

Note that in case when $t_2=\bar t_2$ for $f$ to be real it is
necessary that $G=\bar H$, i.e. $f=F+2$Re$G$.


\section*{\large\bf \quad References.}
\begin{enumerate}
\bibitem{witten}
     Witten E 1981 {\it Nucl. Phys.  }   {\bf B188}   513\\
    Lahiri A  Roy P K and Bagchi B  1990 {\it Int. J. Mod. Phys.}
    {\bf A5}    1383\\
    Cooper F   Khare A  and Sukhatme U  1995    {\it Phys. Rep.}  {\bf 25}
       268
\bibitem{Vinet}
D'Hoker E and Vinet L 1985 {\it Commun. Math. Phys.} {\bf 97}
\bibitem{Rittenberg}
de Crombrugghe M and Rittenberg V 1983 {\it Ann.Phys.} {\bf 151}
99
 \bibitem{Levai}
Levai G and Cannata F 1999 {\it Journ. Phys} {\bf A32} 3947
\bibitem{abis}
     Andrianov A A  Borisov N V Ioffe M V and Eides M I 1984 {\it Theor. Math. Phys.}
{\bf 61} 965 [transl  from 1984 {\it Teor. Mat. Fiz.} {\bf 61}
17]\\
Andrianov A A  Borisov N V Ioffe M V and Eides M I  1985 {\it
Phys. Lett.} {\bf A109} 143
\bibitem{abif1}
     Andrianov A A  Borisov N V and Ioffe M V 1984
{\it Phys. Lett.} {\bf A105} 19 \\
Andrianov A  A  Borisov N  V  and Ioffe M  V 1985 {\it Theor.
Math. Phys.} {\bf 61} 1078  [transl  from 1984 {\it Teor. Mat.
Fiz.} {\bf 61} 183]
\bibitem{ai}
     Andrianov A A and M Ioffe M V 1988 {\it Phys. Lett.}    {\bf B205}
      507
\bibitem{Ahi}
 Berestetskii V  B  Lifshitz E  M  Pitayevskii L  M 1982 {\it Quantum
Electrodynamics} (Oxford: Pergamon)
\bibitem{Clark}
    Clark T E Love S T and Nowling S R 2000 {\it Mod. Phys. Lett.} {\bf A15} 2105
\bibitem{Voronin}
    Voronin A I 1990 {\it Phys. Rev.} {\bf A43} 29\\
    de Lima Rodrigues R Bezerra V B and Vaidya A N 2001 {\it Phys. Lett.}
{\bf A287} 45
 \bibitem{Klish}
    Klishevich S M and Plyushchay M S 2001 {\it Nucl. Phys.} {\bf B616} 403
\bibitem{Nikitin1}
 Niederle J and Nikitin A 1999 {\it Jour. Math. Phys.} {\bf 40} 1280
\bibitem{Nikitin2}
 Nikitin A 1999 {\it Int. Jour. Mod. Phys.} {\bf 14} 885
\bibitem{cooper2}
Cooper F   Khare A Musto R and Wipf A  1988    {\it Ann. Phys.}  {\bf 187} 1

\end{enumerate}
\end{document}